# MODELING THE LIGHTCURVES OF TYPE Ia SUPERNOVAE[1]

W. D. VACCA
*Institute for Astronomy*
*2680 Woodlawn Dr., Honolulu, HI 96822, USA*

AND

B. LEIBUNDGUT
*ESO*
*Karl-Schwarzschild-Strasse 2, D-85748 Garching, Germany*

**Abstract.**
In order to investigate non-uniformity in the luminosity evolution of Type Ia supernovae, we fit the lightcurves with a multi-parameter empirical model. The model provides a quantitative method of analyzing the lightcurves of Type I supernovae and a convenient and continuous representation of the photometric data. Using photometry of the well-observed event SN 1994D we demonstrate the application of the model and construct a bolometric lightcurve. The rise time of the bolometric light curve can be estimated and compared to current theoretical models. By applying this model to a large number of SN Ia lightcurves in various filters, we can explore the variations among individual events and derive systematic correlations. We find that the initial decline rate after maximum light correlates with the decline rate at late phases. Such correlations can be used to constrain theoretical models.

## 1. Introduction

The lightcurves of Type Ia supernovae (SNe Ia) reveal the temporal evolution of the radiative flux in photometric passbands. The apparent uniformity of the lightcurves of SNe Ia was recognized in the very earliest studies (Minkowski 1961, Zwicky 1965). Barbon, Ciatti, & Rosino (1973) noted the similarity of the *B* band lightcurves of 38 SN I and generated a composite lightcurve. Doggett & Branch (1985) used these and other data to synthesize a mean Type I lightcurve; they also constructed average lightcurves for Type II SNe. Leibundgut (1988) computed average magnitudes, as a function of time past maximum light, from the photometric data for the best observed SN I events and constructed general "template" lightcurves for Type Ia and Ib supernovae in several filter bands. Covering the phase from about 5 days before until 110 days after maximum light, these templates have been widely used and provide a means of estimating the time and magnitudes of maximum light for events with poorly sampled data. That such general descriptions of the luminosity evolution of SNe Ia provide fairly good representations of the observed data (Leibundgut et al. 1991a) indicates a substantial degree of homogeneity among the members of this SN

---

[1] To appear in the Proceedings of the NATO Advanced Study Institute on *Thermonuclear Supernovae*, eds. R. Canal, P. Ruiz-Lapuente, & J. Isern, (Dordrecht: Kluwer Academic Publishers)

subclass. This realization, together with the observed similarity of the spectra, has led to the adoption of SNe Ia as standard candles for cosmological distance measurements and the derivation of the cosmological expansion parameter $H_0$ (e.g. Kowal 1968, Tammann & Leibundgut 1990, Hamuy et al. 1995, Vaughan et al. 1995).

However, as more and better data on Type Ia SNe have become available with improved detectors, dedicated searches, and extensive follow-up observations, it has become clear that variations among SNe Ia do indeed exist (Phillips et al. 1987, Frogel et al. 1987, Filippenko et al. 1992, Leibundgut et al. 1993, Phillips 1993, Maza et al. 1994, Hamuy et al. 1995). The lightcurves of SN Ia are *not* identical and some supernovae have displayed pronounced deviations from the means represented by the templates. The most prominent examples are SN 1986G (Phillips et al. 1987, Leibundgut 1988), SN 1991T (Phillips et al. 1992), and SN 1991bg (Filippenko et al. 1992, Leibundgut et al. 1993). Phillips (1993) demonstrated that the difference (in magnitudes) between the brightness at the peak and that at 15 days after the maximum varies considerably among SNe Ia events. In addition, he found a correlation between this difference ($\Delta m_{15}$) and the absolute magnitude reached at maximum. This has far-reaching implications for the use of SNe Ia as standard candles. Variations in luminosities and decline rates have also been found by Hamuy et al. (1993, 1995) using the extensive and homogeneous sample of SNe Ia lightcurves obtained as part of the Calan/CTIO supernova search. A correlation between $\Delta m_{15}$ and $M_{max}$ is also observed in this sample, although the slope of the relation is slightly different from that determined by Phillips (1993). An independent analysis, which uses the SN Ia template as a basis function to which time-dependent corrections are applied to account for variations among the observed lightcurves, yields a similar result for the same data set (Riess et al. 1995; their Figure 1).

In order to investigate the differences in SNe Ia lightcurves, a quantitative description of their shapes is needed. We have adopted a simple analytic expression, incorporating a few free parameters, and used it to fit the photometric data of all known SNe Ia (Vacca & Leibundgut 1996). While not based on any particular physical model, this purely empirical form accounts for the various phases of SNe I lightcurves and provides a continuous representation of the data. The fitting procedure allows us to obtain objective and quantitative measurements of a set of parameters characterizing the lightcurves of individual supernovae, including the maximum magnitude, the time of maximum, and the decline rates at various phases. In addition, an estimate of the error in the observed photometry or the errors in the fitted parameters can be obtained. While several recent analyses of SNe Ia lightcurves are based on a set of basic shapes and deviations therefrom (Hamuy et al. 1995, Riess et al. 1995), our method assumes only that we can approximate lightcurves with a smooth function of the adopted form. Thus, we can independently test the assumption of a continuous sequence of lightcurve shapes in the analysis of photometric data. Here we briefly describe our model and outline some of its applications. A detailed publication of the fits to the observed SN lightcurve data will be forthcoming (Vacca & Leibundgut 1996). We are currently improving the model and extending its application to additional types of SNe.

## 2. The Empirical Model

Leibundgut et al. (1991a) presented an atlas of all SNe I lightcurves known as of 1984. The shapes of both the observed lightcurves and the template curves suggest a simple analytical form for Type I lightcurves. We model the evolution of the observed brightness (in magnitudes in a given filter band) as a function of time with a Gaussian atop a linear decay. The late-time decline is fitted by the line while the peak phase is represented by the Gaussian. A second Gaussian is introduced to fit the secondary bump or maximum which is observed in the $V$, $R$, and $I$ lightcurves (see below). To account for the rising branch of the lightcurves we multiply this function by an expression which rises sharply and approaches unity near the peak of the first Gaussian. This simple approximation, with only a few free parameters, yields a surprisingly accurate representation of observed lightcurves in several filter passbands. In order to test the model, we fitted the $U$, $B$, and $V$ template lightcurves tabulated by Leibundgut (1988, see also Cadonau et al. 1985) using a least-squares technique. We found that the model can easily and accurately reproduce the templates to within about 0.05 mag in the $U$ and $V$ filters and within about 0.01 mag in the $B$ filter over the entire temporal range.

One of the most useful aspects of a continuous model of the lightcurve shapes is that additional interesting quantities and parameters can be objectively derived from it. In particular, we derive the time of maximum brightness, $T_{max}$, the peak apparent magnitude, $m_{max}$, the magnitude difference between the peak and 15 days after the peak, $\Delta m_{15}$, the slope of the intial decline after maximum, $s_1$, the slope of the late-time decline, $s_2$, and the time and magnitude difference between the peak and the "inflection point" in the lightcurve, $\Delta t_{infl.}$ and $\Delta m_{infl.}$, respectively. This inflection point is defined to be the time at which the contribution of the Gaussian to the total apparent magnitude is equal to 0.10 mag. The values of these parameters derived from our fits to the templates in each photometric band are given in Table 1. The overall systematic accuracy of the fit procedure can be judged from the entries in Table 1; the templates were specifically constructed to have $T_{max,B} = 0.0$ and $B_{max} = 0.0$, $T_{max,V} = 2.5$ and $V_{max} = -0.02$, and $T_{max,U} = -2.8$ and $U_{max} = -0.08$. Clearly, this fitting technique can provide a fairly accurate representation of the observed lightcurves.

TABLE 1. Derived Parameters for SN Ia Lightcurves

| Band | $T_{max}$ (day) | $m_{max}$ (mag) | $s_1$ (mag/day) | $\Delta m_{15}$ (mag) | $s_2$ (mag/day) | $\Delta t_{infl.}$ (days) | $\Delta m_{infl.}$ mag |
|---|---|---|---|---|---|---|---|
| | | | Templates | | | | |
| $U$ | $-2.76$ | $-0.05$ | 0.128 | 1.23 | 0.025 | 36.8 | 3.20 |
| $B$ | 0.06 | 0.00 | 0.115 | 1.11 | 0.016 | 37.3 | 2.87 |
| $V$ | 2.53 | 0.03 | 0.070 | 0.61 | 0.026 | 34.4 | 1.74 |
| | | | SN 1981B | | | | |
| $U$ | $4670.7^a$ | 11.83 | 0.140 | 1.36 | 0.035 | 33.1 | 3.19 |
| $B$ | 4672.9 | 12.05 | 0.113 | 1.14 | 0.010 | 41.6 | 3.08 |
| $V$ | 4673.4 | 11.96 | 0.067 | 0.59 | 0.023 | 42.4 | 2.12 |
| | | | SN 1989B | | | | |
| $U$ | $7564.4^a$ | 12.37 | 0.154 | 1.65 | 0.026 | 29.1 | 3.03 |
| $B$ | 7565.2 | 12.32 | 0.127 | 1.28 | 0.015 | 34.9 | 2.93 |
| $V$ | 7566.0 | 11.93 | 0.065 | 0.60 | 0.024 | 42.3 | 2.09 |
| | | | SN 1994D | | | | |
| $U$ | $9432.2^a$ | 11.24 | 0.178 | 1.94 | 0.027 | 29.4 | 3.48 |
| $B$ | 9432.9 | 11.85 | 0.139 | 1.47 | 0.017 | 33.8 | 3.12 |
| $V$ | 9433.4 | 11.90 | 0.054 | 0.83 | 0.031 | 27.4 | 1.71 |

$^a$ JD $-2440000$

By performing a least-squares fit of a parametric model to the lightcurve data (as opposed to simply connecting the scattered data points with a spline, for example), spurious fluctuations due to random observational errors are minimized, errors on all derived quantities can be determined, and an estimate of the goodness-of-fit of the model to the data points can be obtained. Furthermore, the fits yield a far more objective means of quantifying the parameters describing the lightcurves than the templates can provide. As an example we list the lightcurve parameters found for three well observed SNe Ia in Table 1. For these SNe, the intervals between the times of maximum light in the different filters are smaller than those adopted in the templates. In particular, the fits yield times of maximum in $V$ which are within a day of those in $B$ for all three supernovae. Furthermore, the values of the $s_1$, and the decline parameters $\Delta m_{15}$ and $\Delta m_{infl}$ are found to be clearly correlated with filter wavelength, decreasing as the wavelength increases.

The residuals from the fits also provide a means of identifying subtle, yet systematic deviations from the mean evolutionary behavior represented by the templates. For example, when fitting the $V$ lightcurves of some well-observed, recent SNe Ia with our single-Gaussian model, we found a small but significant "bump" in the residuals, corresponding to a second peak in the lightcurve data, at about 15 – 20 days after the primary maximum (see also Ford et al. 1993, Suntzeff 1995). This peak is similar to the secondary maxima observed in the $R$ and $I$ lightcurves of SNe Ia in general (Elias et al. 1981). To reproduce this feature of the lightcurves, we incorporated a second Gaussian in the model; the fits to the observed $V$, $R$ and $I$ lightcurves of SNe Ia improve dramatically when this additional component is included.

## 3. Application to SN 1994D

### 3.1. MODEL FITS

In Figures 1 and 2, we show a representative example of the model fits to the $U$, $B$, $V$, $R$, and $I$ lightcurves of SN 1994D. Three data sets, from Richmond et al. (1995), Patat et al. (1995), and the RGO supernova data archive (e.g., Lewis et al. 1995), have been combined in these figures. The residuals from the fits are shown below each panel. In the $U$, $B$, and $V$ filters, the models fit the observed lightcurve data much better than the templates can over the entire range of observations. Several parameters derived from our fits to the $U$, $B$, and $V$ lightcurve data of SN 1994D are presented in Table 1. (Results for SN 1981B and SN 1989B are also presented in Table 1 for comparison.) Note that immediately after maximum light SN 1994D declined faster than the template in all three filters. The secondary peaks in the $V$, $R$, and $I$ lightcurves are well matched by our model fitting function. This secondary peak is stronger and occurs progressively later at longer wavelengths; the secondary maximum occurs at approximately 20 days ($V$), 22 days ($R$) and 24 days ($I$) after the $B$ maximum. This trend is continued in the $J$ (25 days) and $H$ (29 days) lightcurves of other type Ia supernovae (Elias et al. 1981, Leibundgut 1988). In $K$ the second peak occurs about 20 days after maximum.

### 3.2. A BOLOMETRIC LIGHTCURVE

As shown by Suntzeff (1995), the $U$ through $I$ photometric bands comprise nearly 80% of the total radiation energy emitted by a SN Ia. However, few events have been observed in all of these filter bands. Accurate and extensive observations in all filters do exist for SN 1994D, however, and the construction of a "nearly bolometric" lightcurve for this event is possible. We used our fitting procedure to model the observed lightcurves of SN 1994D in all five photometric bands, and generated a "nearly bolometric" lightcurve. The construction of the bolometric lightcurve is facilitated by our continuous model of the individual filter lightcurves. The resulting bolometric lightcurve is shown at the bottom of Figure 2. The secondary peak observed in the $V$, $R$, and $I$ data carries through to the bolometric lightcurve. A similar feature has been found in the bolometric lightcurve of SN 1992A (Suntzeff 1995). This is an important characteristic which will have to be matched by theory. We then fitted the bolometric lightcurve with our model and determined its parameters; the model can reproduce the bolometric lightcurve to within 0.01 mag over the entire range of observations. The rise time, defined as the difference (in days) between the time of maximum brightness and when the supernova was 30 magnitudes fainter than peak brightness, was found to be $\sim 18$ days for the bolometric lightcurve. The average of the rise time determinations for the individual filter curves is the same, $18.2 \pm 1.4$ days. This is very similar to the value determined for SN 1990N ($\sim 19$ days; Leibundgut et al. 1991b). The comparison between this estimate of the rise time and the predictions from various theoretical models (Khokhlov et al. 1994) yields a very interesting result. All direct detonations and deflagrations yield rise times which are much shorter than that found for SN 1994D. Delayed detonations may have rise times as slow as 15 days, but only (unrealistic) detonations with a damping envelope have rise times as long as that observed for SN 1994D. The relatively slow rise times of SN 1990N and SN 1994D provide strong constraints on the theories of supernova explosions.

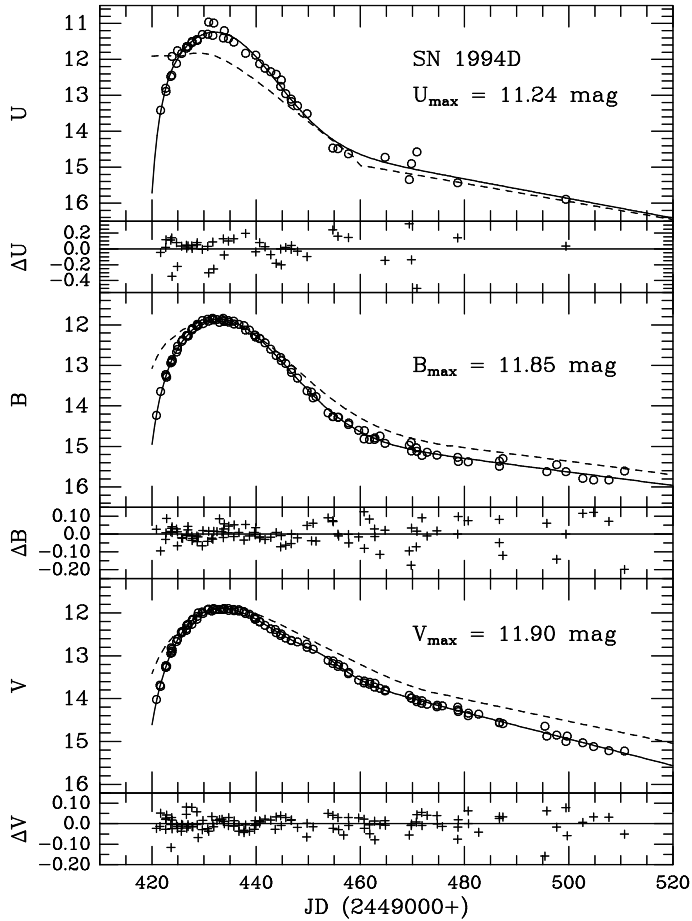

*Figure 1.* The $U$, $B$, and $V$ lightcurves of SN 1994D. The circles are the observed data points. The solid lines are our best fits; the dashed lines are the templates of Leibundgut (1988) normalized to the time and apparent magnitude at maximum in $B$, as given by Richmond et al. (1995). The residuals of the fit are shown below each panel.

## 4. Statistics

The primary reason for adopting an analytic model to parametrize the lightcurves of SNe Ia was to investigate the uniformity of the lightcurves. The results provided by our fits allow us to determine quantitative parameters for many supernovae (most of which have sparsely sampled data), which can then be analysed with statistical techniques in order to examine the variation among SNe Ia lightcurves and gain some insight into the general nature of the events. Using the results of our fits, we can also search for correlations among a large number of lightcurve shape parameters.

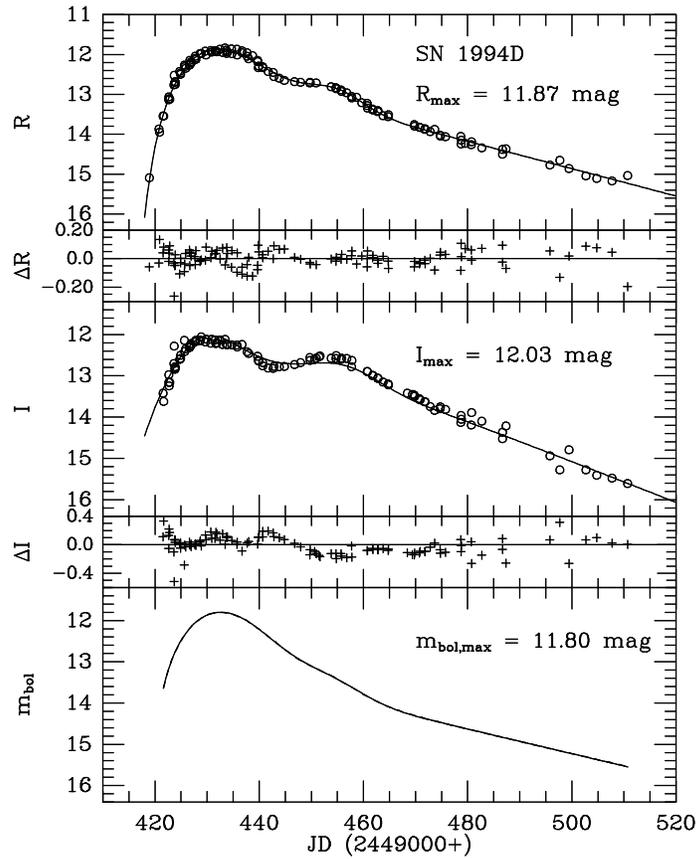

Figure 2. The $R$, $I$, and bolometric lightcurves of SN 1994D. The circles are the observed data points. The solid lines are our best fits. The residuals of the fit are shown below each panel.

We have fitted the lightcurves of all SNe I and SNe Ia with adequate photometric data (Vacca & Leibundgut 1996). A sample of our findings is shown in Figure 3, where we plot the values of $\Delta m_{15}$ and the initial decline rate $s_1$ against those of the late-time decline rate $s_2$ for the $B$ and $V$ filter bands. We confirm the range of $\Delta m_{15}$ found by Phillips (1993). We also find a substantial range in the values of $s_1$ and $s_2$. Since the late phase of the lightcurve is powered by the thermalization of the $\gamma$–rays from the radioactive source, the time scale of which depends only on the column density (Leibundgut & Pinto 1992), this variation in $s_2$ indicates differences in the combination of ejecta mass and explosion energy among the individual SNe. Although there is considerable scatter, trends appear to be present in all four plots. In addition, we find that $\Delta m_{15}$ and $s_1$ are extremely tightly correlated. Such correlations make it possible to determine $\Delta m_{15}$ and the luminosity correction even for those SNe Ia discovered several days after maximum.

## 5. The Color-Color Plot

With our model fits to the $U$, $B$, and $V$ lightcurves we can also follow the continuous evolution of SNe Ia in the color-color diagram. The evolution in the $U - B$ vs. $B - V$ plane for three well-observed SNe is shown in Figure 4. Note the characteristic shape of the curves. The loops occur approximately at the time of the inflection in the lightcurves. Because of this characterstic shape, it is possible that the color-color diagram for SNe Ia can provide rough estimates of the reddening toward various objects. For example, if we assume the two events were nearly identical, then Figure 4 indicates that SN 1989B experienced approximately one magnitude of reddening in $V$ relative to SN 1981B, a result which is in agreement with the analysis of Wells et al. (1994). It is also clear that SN 1994D was unusually bright in $U$ (see also Patat et al. 1995 and Richmond et al. 1995). Furthermore, it can be seen that between about 25 and 40 days past maximum light, SNe Ia have colors similar to those of fairly cool blackbodies ($T_{eff} = 4000 - 5000$ K). Moreover, the spectra of SNe Ia at this time appear to be very similar (e.g., Leibundgut et al. 1993). These results suggest that the inflection point may represent a phase in the lightcurve more appropriate than the maximum for simple reddening determinations. After 40 days the spectra are dominated by emission lines, rather than continuum emission, and a comparison with blackbodies is not physically meaningful.

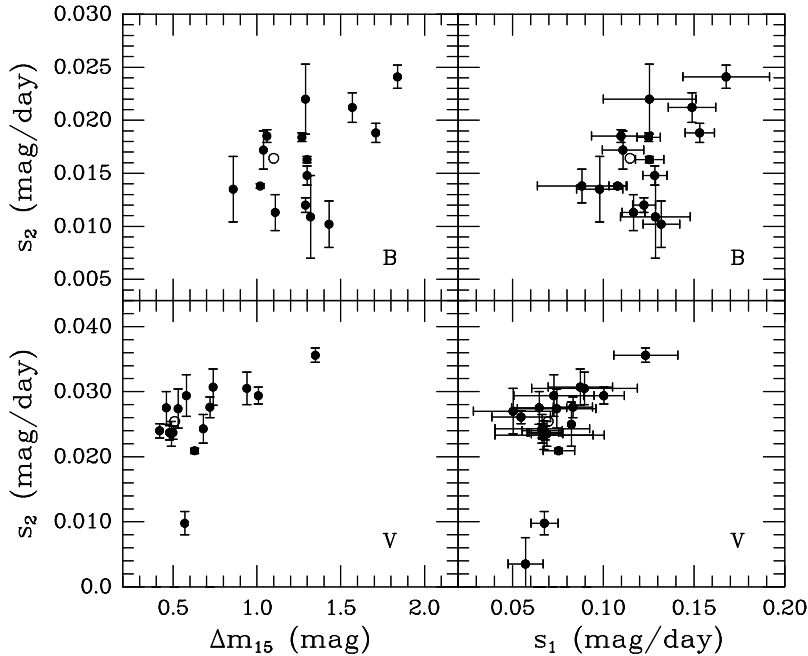

*Figure 3.* The correlation between $\Delta m_{15}$ and $s_2$, and between $s_1$ and $s_2$, as derived from our model fits to $B$ and $V$ band photometric data for SNe Ia. The open circles denote values derived from the templates of Leibundgut (1988).

## 6. Conclusions

The assumption that SNe Ia form a homogenous set of events is the basis for the use of these objects as cosmological standard candles. However, much evidence has become available over the last few years which invalidates this assumption. We have developed an objective way to quantify variations in the lightcurves of SNe Ia by a simple fitting technique. Through an accurate description of the irregularly sampled data an objective comparison of individual events can be performed. By investigating the observed variations in the lightcurves we will be able to constrain the physical models of SNe Ia and also establish the degree of suitability of these events for cosmological applications. For supernovae with adequate observations we can also simplify the construction of bolometric lightcurves, thus forming a link between theory and observations.

We have presented some preliminary results of our fitting procedure. We detect a secondary bump in the bolometric lightcurve of SN 1994D, similar to that found in the bolometric lightcurve of SN 1992A. The rise time of the bolometric curve excludes models which exhibit sharp increases to maximum light. The color-color plot may become an independent way to find reddening towards supernovae, if it can be established that SNe Ia cool to a similar temperature after the peak phase. From a sample of available lightcurves of SNe Ia we find a correlation between the early decline rate (or $\Delta m_{15}$) and the late decline rate. Such a correlation is an implicit assumption in the fitting techniques employed elsewhere. The range of late-time decline rates is surprisingly large and directly reflects differences in the ejecta mass, the explosion energies, or a combination of the two among individual events. Clearly the lightcurves of SNe Ia are providing a wealth of information that we are only beginning to extract.

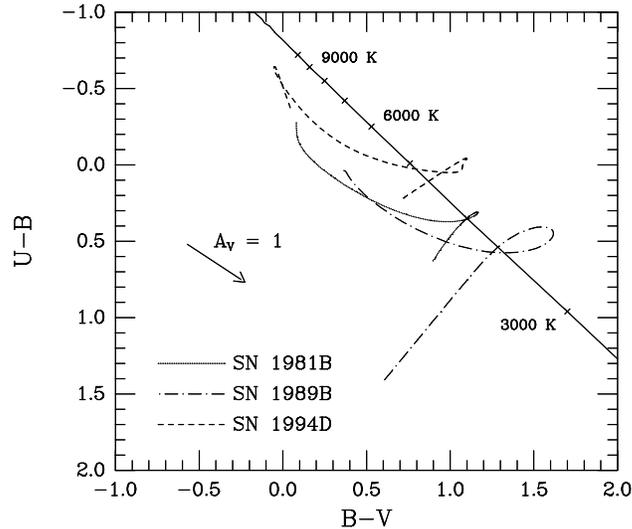

*Figure 4.* The evolution of three SNe Ia in the color-color diagram. The solid diagonal line denotes the location of blackbodies of various effective temperatures. The reddening vector is also shown.